\documentclass[10pt,conference,a4paper,hidelinks]{IEEEtran}

\usepackage[utf8]{inputenc}

\usepackage{times}

\usepackage{graphicx}
\usepackage{subfigure}
\usepackage{amsmath}
\usepackage{hyperref}
\usepackage{academicons}
\usepackage{xcolor}
\usepackage{amsmath,amssymb}
\usepackage{cuted}
\DeclareGraphicsExtensions{.png,.eps,.ps,.pdf}

\usepackage{url}

\usepackage[english]{babel}

\begin{document}

\title{A Security and Trust Framework for \\Decentralized 5G Marketplaces}

\author{\IEEEauthorblockN{José María Jorquera Valero, Manuel Gil Pérez, and Gregorio Martínez Pérez}
\IEEEauthorblockA{Faculty of Computer Science, University of Murcia, 30100 Murcia, Spain\\
Email: \{josemaria.jorquera, mgilperez, gregorio\}@um.es}
}

\maketitle

\begin{abstract}
5G networks intend to cover user demands through multi-party collaborations in a secure and trustworthy manner. To this end, marketplaces play a pivotal role as enablers for network service consumers and infrastructure providers to offer, negotiate, and purchase 5G resources and services. Nevertheless, marketplaces often do not ensure trustworthy networking by analyzing the security and trust of their members and offers. This paper presents a security and trust framework to enable the selection of reliable third-party providers based on their history and reputation. In addition, it also introduces a reward and punishment mechanism to continuously update trust scores according to security events. Finally, we showcase a real use case in which the security and trust framework is being applied.
\end{abstract}

\begin{IEEEkeywords}
Security, Trust Management, 5G, Trustworthy Relationships
\end{IEEEkeywords}

{\bf Tipo de contribución:}  {\it   Investigación en desarrollo}

\section{Introduction}
The fifth generation of mobile networks (5G) entails a continuously increasing figure of interconnected end-users as well as relationships among entities. Thereby, 5G resource, service, and infrastructure providers are constantly working on innovative solutions to deal with the huge demand from tenants and users for data capacity, bandwidth, network coverage, and latency. One of the most promising approaches is \textit{decentralized marketplaces}. A marketplace is normally described as a centralized or decentralized repository in which thousand of offers are advertised by resource and infrastructure providers. The trading of such resources and services is carried out by consumers who need to cover dynamic requirements and high Quality-of-Service (QoS); for instance, high speed, low energy consumption, and automatically satisfy user demands. In this context, marketplaces boost the generation of service chains across operators with security and trustworthiness \cite{fernandez2021multi}.


Conventionally, trust has been employed as a mechanism capable of determining the trustworthiness level that a trustor has in a trustee. In this regard, there are multiple models for determining a trust level such as PKI-based, role-based, or reputation-based, among others. The latter is one of the most considered models, since it not only allows the trust to be calculated based on entity's behavior in previous interactions but also allows gathering feedback from reliable recommenders who also interacted with the entity. Yet, reputation-based trust models also open the door to key challenges to be addressed whether they want to be one of the technologies that await to support next-generation solutions (5G and beyond) \cite{valero2022toward}.

Especially, trust models should guarantee a minimum set of features in order to be integrated with promising 5G solutions. First and foremost, trust models need to provide highly dynamic and context-dependence solutions as 5G ecosystems tend to support scenarios where main actors scale and migrate flexibly. In this sense, trust models should continuously collect information from the principal actors involved in the relationship through automatic mechanisms triggered by events, rules, time, etc. By means of these mechanisms, the trust score can be rapidly adapted in real-time, and in consequence, adjust the participating entities, if necessary. Besides, trust models should ensure reliable end-to-end establishments, therefore any intermediate entities must be analyzed. Another pivotal aspect to be covered is the automated management following a zero-touch approach. On the one hand, trust models should leverage tools that enable the automatization of network and service management via high-level policies, and artificial intelligence algorithms. On the other hand, they should empower a flexible integration with other 5G essential services and their workflows. Lastly, trust models should fulfill the zero trust principle \cite{rose2020zero}. It attempts not to attribute implicit trust to an entity regardless of whether the trustor and trustee had previously a relationship or whether both entities belong to the same domain. Therefore, trust models would avoid utilizing an outdated trust value without being updated at a given time as well as assign trust-by-default to entities under our reliable zone.

\begin{figure*}
    \centering
    \includegraphics[width=0.88\linewidth]{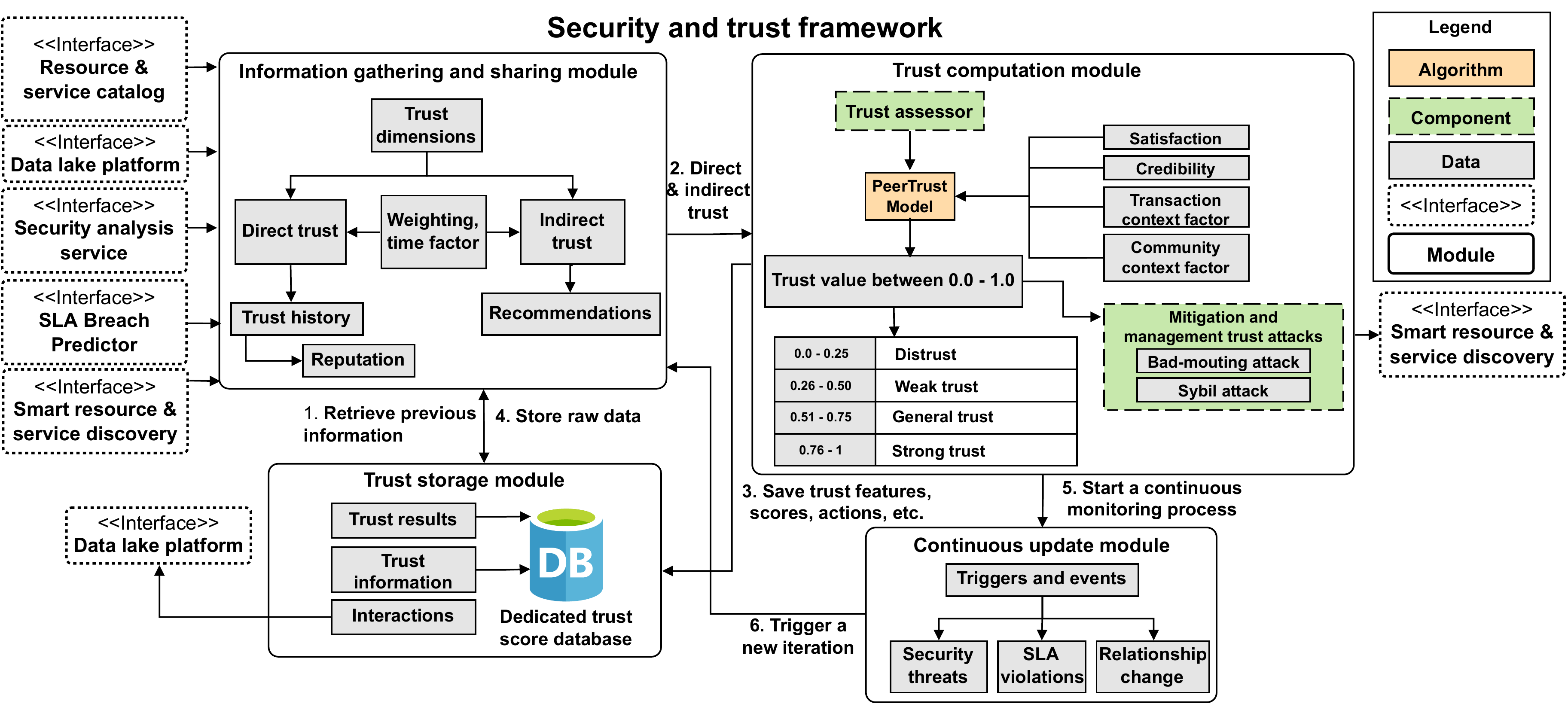}
    \caption{Modular architecture of the proposed security and trust framework}
    \label{fig:reputation_trust_framework_design}
\end{figure*}

Despite the progress of literature, more efforts are needed to enhance the trust and reputation models in 5G scenarios as prior trust models did not contemplate all requirements set out above. Furthermore, the trust concept also includes security aspects, therefore considering security-related features for profiling actor behavior would help to detect feasible threats as well as to have a broader view when determining trust. In this regard, this paper presents a security and trust framework capable of ensuring a trustworthy network inside of a decentralized marketplace. In particular, such a framework enables consumers to establish trustworthy end-to-end connections across domains, at the same time it covers the aforementioned challenges. To this end, the security and trust framework analyzes a set of product offers, published by resource and infrastructure providers in a decentralized marketplace, in order to predict trust scores from previous interactions as well as recommendations from trustworthy third parties. It should be pointed out that such a framework is being developed in the 5GZORRO H2020 European project \cite{carrozzo2020ai}, which enables a secure, flexible, and distributed multi-stakeholder combination and composition of resources and services in 5G networks. Additionally, the security and trust framework also introduces a reward and punishment mechanism to continuously update trust levels from multiple security events obtained in real-time. Finally, the paper introduces a real use case to demonstrate how the framework is integrated with other 5G services and helps to perform a smart resource and service discovery.

The remainder of this paper is structured as follows. Section \ref{sec:framework} presents an overview of our security and trust framework, detailing the principal modules and the main actions to be performed. Section \ref{sec:use_case} introduces the use case in which our framework is being applied. Finally, Section \ref{sec:conclusion} expounds some conclusions an
open perspectives for future work.

\section{The Security and Trust Framework}
\label{sec:framework}

This section describes a modular architecture of the proposed security and trust framework which enables to cater to a trustworthy ecosystem among multiple domains and operators. As Fig. \ref{fig:reputation_trust_framework_design} sketches, the security and trust framework is made up of four principal modules: the \textit{information gathering and sharing}, \textit{trust computation}, \textit{trust storage}, and \textit{continuous updates} modules. In this vein, this section is going to thoroughly explain them as well as the capital actions carries out. 

\subsection{Information gathering and sharing module}

One of the uppermost important steps in a security and trust framework is the collection of meaningful features to afterward evaluate entity behavior. In this sense, it is pivotal to find out a set of information sources from which the trustor can retrieve information to ascertain the credibility and effectiveness of the assigned trust scores. Due to the fact that this framework is designed bearing in mind the 5GZORRO ecosystem, we have selected four principal software modules from which we will infer features (see interfaces in Fig. \ref{fig:reputation_trust_framework_design}). On one side, we consider the Catalog module \cite{D2_3} which stores the Product Offerings (POs) published by the stakeholders registered in the 5GZORRO system. Note that there are seven types of product offers defined in the Catalog: radio access network (RAN), spectrum, virtual/container network function (VNF, CNF), slice, network service, cloud, and edge. By means of this dimension, the security and trust framework is capable of deriving statistic features such as the number of POs available per provider, per location, per type of resource, or what POs were available but are not currently, among other features. On another side, we contemplate the Data Lake platform \cite{D2_3} which stores all the interactions performed by stakeholders in the past. Thereby, a consumer would be able to know what stakeholders previously interacted with a specific resource and service provider and, if necessary, requests recommendations (indirect trust) about the provider. This information source plays an important pillar for the security and trust framework as this is based on a reputation model. Thence, recommendations from third parties are considered one of the two types of trust (direct and indirect trust) to compute a final trust score. Concerning direct trust, it is formulated from the previous interactions with the target stakeholder as well as new features inferred from Catalog.

In addition, the security and trust framework also gathers information from the Security Analysis Service (SAS) \cite{D2_3} which is instantiated once an end-to-end relationship is set up. Hence, it is not leveraged when the framework is analyzing a set of POs but when a final PO is selected. Through the SAS component, the framework can detect security threats or misbehavior in the network traffic as the SAS monitors, through the Zeek tool \cite{zeek}, network packets sent by consumers and providers. Moreover, the SAS component has activated several security policy rules that allow triggering alerts when unusual behaviors happen in real-time. Last but not least, the security and trust framework also recaps information from the Intelligent Service Level Agreement Breach Predictor (ISBP) module \cite{D2_3}. In this case, the framework leverages events published by the ISBP so as to predict the provider's reputation and how it can be affected based on current events such as Service Level Agreement (SLA) violations. 



\subsection{Trust computation module}

Once the framework collects raw data from the aforementioned information sources, it is time to determine a trust score per each PO to be analyzed. Note that this step is performed regardless of whether stakeholders have a previous trust relationship for a while or whether they belong to the same domain (zero trust). Since the 5GZORRO ecosystem boosts a decentralized philosophy where peer-to-peer communications are carried out, we have selected the PeerTrust model \cite{peerTrust} as a statistic algorithm to prognosticate a trust score. The PeerTrust model is an algorithm based on the reputation that has been utilized in decentralized scenarios for ages. Concretely, the PeerTrust model determines trust scores from four primary pillars: Satisfaction (S), Credibility (Cr), Transaction Factor (TF), and Community Factor (CF), as depicted in Eq. (\ref{eq:general_peerTrust}). Note that the PeerTrust model only introduces the above-mentioned generic pillars but each researcher should find out how each one will be formulated. Thus, our security and trust framework has carried out an adapted PeerTrust model.

\begin{equation}
     T(u) = \alpha \cdot \displaystyle\sum_{i=1}^{I(u)} S(u,i) \cdot Cr(p(u,i)) \cdot TF(u,i) + \beta \cdot CF(u),
     \label{eq:general_peerTrust}
\end{equation}

Since each pillar is composed of multiple equations, this paper does not display the 19 equations formulated to cover all pillars. Instead, we broadly explain below how each pillar is computed as well as presenting some of the utmost important equations. In the case of \textit{Satisfaction} (S), the framework predicts the provider's satisfaction and the offer's satisfaction. The main difference between them is the former considers all assets linked to a provider whereas the latter only envisages a given type of PO. Thence, the provider's and the offer's satisfactions are formulated from the reputation, a set of recommendations about the target, and the last trust score we have for each recommender in the previous set. When it comes to \textit{Credibility} (Cr), the adapted PeerTrust model leverages a Personalized Similarity Metric (PSM) through which we can discover how similar two stakeholders are when they are evaluating the same targets. Thus, the PSM allows us to know the distance of credibility of a set of evaluated stakeholders as well as contrast their opinions.

Additionally, the adapted PeerTrust model also takes into account two context factors: the \textit{Transaction Factor} (TF) and the \textit{Community Factor} (CF). Concerning the TF, it intends to predict a trust value linked to the current interaction, with a particular stakeholder or product offer, from the number of feedbacks published in the Data Lake from different time windows. As a result, a higher number of feedbacks published in the Data Lake, a higher number of recommenders to be contemplated for finally computing a stable reputation. The TF then rewards stakeholders who publish their interactions with others in the Data Lake since it spurs future stakeholders to look into the Data Lake, request recommendations to other stakeholders, and grow the community. Instead, the CF intends to assess the stakeholder participation within the community to give greater or lesser weight to their recommendations over time. To this end, the CF collects multiple feedback from $j$-th trustworthy opinions on a target stakeholder $u$ for computation, see Eq. (\ref{eq:CF_peerTrust}), and notices dishonest recommendations as rated by the confidence from recommenders (CR) on $u$. By CF, we can thus detect conventional attacks in trust models such as bad-mouthing attacks.




\begin{equation}
    CF(u) = \frac{\frac {R(u)}{I(u)} +  \displaystyle\bigoplus_{j=1}^{n} CR(v,j,u) \cdot Inf(v,j)}{2},
    \label{eq:CF_peerTrust}
\end{equation}

where $R(u)$ denotes the number of published recommendations and $I(u)$ the total number of interactions of $u$; and $Inf(v,j)$ the recommender's influence on all contemplated recommendations $Rec(j,u)$. Furthermore, Eq. (\ref{eq:CR_peerTrust}) shows how to measure the confidence of a $j$-th recommender on $u$, in accordance to the trust on that recommender $T(v,j)$ and the recommendation trust $RT(v,j)$.

\begin{equation}
    CR(v,j,u) = \alpha \cdot T(v,j) + (1 - \alpha) \cdot \big( RT(v, j) \cdot Rec(j,u) \big),
    \label{eq:CR_peerTrust}
\end{equation}

Moreover, the CF also measures the number of interactions that a specific stakeholder had in the community through the contribution of services or resources with other stakeholders. In the end, multiple recommendations together with the credibility of the recommender are contemplated through an aggregation function to achieve the general reputation of the community about a target stakeholder.


\subsection{Trust storage module}

Another capital step of the security and trust framework is the storage of non-public data, raw data, and new interactions between stakeholders. In this vein, the framework employs two kinds of storage sources. On the one hand, the security and trust framework manages both raw data and all features inferred from them which are stored in a private database per domain. In this case, we are handling information which must not be public to other stakeholders. On the other side of the coin, the 5GZORRO ecosystem takes advantage of a Data Lake platform that allows users to push any type of information. By means of the Data Lake, the framework informs other stakeholders about the interactions. Concretely, after establishing a trust relationship the trustor pushes a new object into the Data Lake with information on the parties involved, the start date, the number of interactions between them, etc. In this way, newcomers can look at the Data Lake and identify what trustworthy recommenders had interaction with a particular target, and consequently, request their feedback about their relationships. In addition to that, the Data Lake introduces pivotal characteristics such as decentralization, a long-term reputation reflection, and security.

\subsection{Continuous update module}
Parallel to the trust storage module, the next step is to continuously monitor an ongoing trust relationship in order to adapt a previous trust score to the events occurring in real-time. Hence, this module plays an essential role since it not only enables early threat identification but also enhances the security capabilities of the framework through dynamicity, context-dependence, and end-to-end automatization. This module includes a reward and punishment mechanism which allows collecting security-based network monitoring events so as to increase or dwindle a trust score. This mechanism follows a time-driven approach so we need to declare pre-established time windows in which the previous trust score will be updated using the network monitoring events. To update the previous trust scores, the security and trust framework makes use of four main log files gathered by Zeek: \textit{conn.log}, that gathers the tracking of general information regarding TCP, UDP, and ICMP traffics; \textit{notice.log}, that collects likely monitoring events which are potentially odd; \textit{weird.log}, that takes action when unusual or exceptional activity appears; and \textit{stat.log}, that obtains memory, packet, and log statistics. These four weighted logs are shown in Eq. (\ref{eq:dimension_RP}) to compute reward and punishment (RP) values.

\begin{equation}
    \begin{split}
        RP(v,u) = \alpha\cdot Conn(v,u) + \beta\cdot Notice(v,u) +\\ \cdot Weird (v,u) + \phi\cdot Stat(v,u)
    \end{split}
    \label{eq:dimension_RP}
\end{equation}


Finally, depending on the RP value obtained for the current time window, the framework will update the prior trust value ($O_{ts}$) to a new trust score ($N_{ts}$) accordingly, as given in Eq. (\ref{eq:RW_peerTrust}). Note that a value close to the extremes will result in a larger increase or decrease than if the RP value is around 0.5.

\begin{equation}
    \resizebox{.99\hsize}{!}{$
    N_{ts}(v,u) = 
    \begin{cases}
        $$ O_{ts}(v,u) + (RP(v,u) - 0.5) \cdot \frac{\Big(1 - O_{ts}(v,u)\Big)}{10} $$,& \text{if } RP(v,u) \geq 0.5\\
        $$ O_{ts}(v,u) - (0.5 - RP(v,u)) \cdot \frac{\Big(1 - O_{ts}(v,u)\Big)}{10} $$,              & \text{if } RP(v,u) < 0.5
    \end{cases}$}
     \label{eq:RW_peerTrust}
\end{equation}

\section{5GZORRO Use Case}
\label{sec:use_case}

\begin{figure}[h!]
    \centering
    \includegraphics[width=\linewidth]{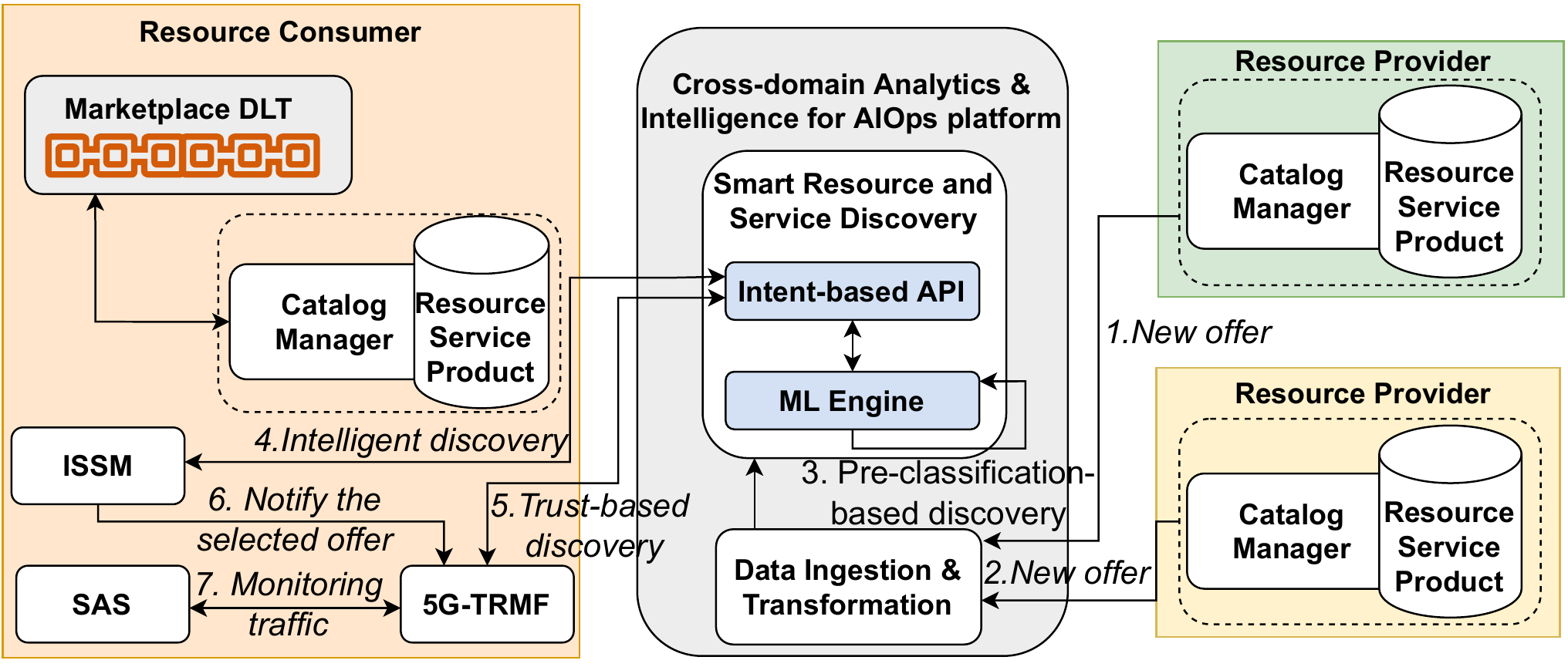}
    \caption{Trust-based resource discovery}
    \label{fig:UC}
\end{figure}

This section describes how the security and trust framework can assist in the process of selecting trustworthy resources and services available in a distributed marketplace (see Fig. \ref{fig:UC}).

Since the generation and registering of offers are not the main focus of our use case, we have simplified the sub-tasks of these steps in order to fully focus on the consumer side. In this case, the consumer detects a lack of resource capabilities to cover the agreed QoS and the dynamic requirements. To overcome the above problems, the consumer decides to extend its current capabilities to satisfy a signed contract and its SLA. At this point, the 5GZORRO marketplace \cite{D2_3} aims at facilitating multi-party collaboration in dynamic 5G environments where consumers and service providers often need to employ third-party resources. As a result, resource providers usually offer their resources available by promoting them through the 5GZORRO marketplace. In general terms, the distributed marketplace allows creating and acquiring product offers that cover a variety of telco digital assets.

To cope with this, the resource consumer leverages the Intelligent Slice and Service Manager (ISSM) \cite{D2_3} to request new resource capabilities as well as looking at the available offers in the . In this sense, two different discovery processes take place via the Smart Resource and Service Discovery (SRSD) \cite{D2_3}. On the one hand, the SRSD conducts a pre-classification of new POs based on imperative constraints such as category, geographic location, price, etc. On the other hand, the ISSM launches an intent-based discovery request employing priorities such as performance, proximity, cost, etc. Afterward, the SRSD determines trust scores and sends back to the ISSM a ranked list of product offers based on the reputation of both the provider and its type of offer. Therefore, the security and trust framework contributes directly to the intelligent discovery process classifying the most trustworthy POs which previously matched the intent-based criteria.

Afterward, the ISSM carries out additional optimization steps to find out the best offer from the list ranked by trust scores. As a next step, the ISSM informs the security and trust framework, called 5G-enabled Trust and Reputation Management Framework (5G-TRMF) in the 5GZORRO ecosystem \cite{D2_3}, about the final offer selected by the consumer. To close the lifecycle of trust-based resource discovery, the security and trust framework enables, as a background process, ongoing monitoring of network traffic between the resource consumer and the new third-party via the Security Analysis Service. As shown in Fig \ref{fig:UC}, our framework can perfectly work with other pivotal modules such as the ISSM, Catalog, and SRSD to help consumers to address day-to-day problems such as resource shortages, while respecting automation and ensuring security and trustworthiness throughout the process.

\section{Conclusion}
\label{sec:conclusion}

This paper presents a security and trust framework capable of ensuring a reliable ecosystem where stakeholders can set up trustworthy end-to-end connections across domains. In this vein, a modular architecture of the proposed framework has been explained, as well as the capital steps under each module. At the same time, the framework also introduces a reward and punishment mechanism to continuously update an ongoing trust relationship via security-based monitoring events generated in real-time.

As future work, we will carry out the validation of the current security and trust framework in real testbeds such as 5GBarcelona and 5TONIC. Besides, the continuous update module will be enhanced by adding a novel event-driven mechanism to adjust trust scores from SLA prediction notifications. Additionally, further computation models will be contemplated to analyze their performance and accuracy, as well as enlarging the resilience of the framework to other conventional trust attacks such as on-off or Sybil attacks.

\section*{Acknowledgment}

This work has been supported by the European Commission through 5GZORRO project (grant no. 871533) part of the 5G PPP in Horizon 2020.

\bibliographystyle{IEEEtran}
\bibliography{references}

\end{document}